\documentclass[a4paper,11pt]{article}
\pdfoutput=1 

\usepackage{jcappub} 
 \usepackage{indentfirst}
\usepackage{appendix}

\newcommand{\ck}{c_K}
\newcommand{\tppi}{\ck p_T \Pi_T}

\title{Weighing neutrinos in the scenario of vacuum energy interacting with cold dark matter: application of the parameterized post-Friedmann approach}

\author[a]{Rui-Yun Guo,}
\author[a]{Yun-He Li,}
\author[a]{Jing-Fei Zhang,}
\author[a,b,1]{Xin Zhang\note{Corresponding author.}}
\affiliation[a]{Department of Physics, College of Sciences, Northeastern University, \\Shenyang
110004, China}
\affiliation[b]{Center for High Energy Physics, Peking University, \\Beijing 100080, China}

\emailAdd{guoruiyun110@163.com, liyh19881206@126.com, jfzhang@mail.neu.edu.cn, zhangxin@mail.neu.edu.cn}

\abstract{We constrain the neutrino mass in the scenario of vacuum energy interacting with cold dark matter by using current cosmological observations. To avoid the large-scale instability problem in interacting dark energy models, we employ the parameterized post-Friedmann (PPF) approach to do the calculation of perturbation evolution, for the $Q=\beta H\rho_{\rm c}$ and $Q=\beta H\rho_{\Lambda}$ models. The current observational data sets used in this work include Planck (cosmic microwave background), BSH (baryon acoustic oscillations, type Ia supernovae, and Hubble constant), and LSS (redshift space distortions and weak lensing). According to the constraint results, we find that $\beta>0$ at more than $1\sigma$ level for the $Q=\beta H\rho_{\rm c}$ model, which indicates that cold dark matter decays into vacuum energy; while $\beta=0$ is consistent with the current data at $1\sigma$ level for the $Q=\beta H\rho_{\Lambda}$ model. Taking the $\Lambda$CDM model as a baseline model, we find that a smaller upper limit, $\sum m_{\nu}<0.11$ eV ($2\sigma$), is induced by the latest BAO BOSS DR12 data and the Hubble constant measurement $H_{0} = 73.00 \pm 1.75$ km~s$^{-1}$~Mpc$^{-1}$. For the $Q=\beta H\rho_{\rm c}$ model, we obtain $\sum m_{\nu}<0.20$ eV ($2\sigma$) from Planck+BSH. For the $Q=\beta H\rho_{\Lambda}$ model, $\sum m_{\nu}<0.10$ eV ($2\sigma$) and $\sum m_{\nu}<0.14$ eV ($2\sigma$) are derived from Planck+BSH and Planck+BSH+LSS, respectively. We show that these smaller upper limits on $\sum m_{\nu}$ are affected more or less by the tension between $H_{0}$ and other observational data.}

\begin{document}
\maketitle
\flushbottom

\section{Introduction}
\label{sec1}
Since the discovery of accelerated expansion of the universe, it has been proposed that there is a mysterious ingredient in the universe, called ``dark energy'', which is a component with negative pressure; in other words, it is the source to produce repulsive gravity to drive the acceleration of the universe's expansion (for reviews of cosmic acceleration and dark energy, see e.g.~\cite{Sahni:1999gb,Peebles:2002gy,Padmanabhan:2002ji,Copeland:2006wr,Sahni:2006pa,Frieman:2008sn,Li:2011sd,Bamba:2012cp,Weinberg:2012es,Mortonson:2013zfa}). Dark energy is now a dominant component (it occupies about 70\% of the total energy density) in today's universe. Also, dark energy is undoubtedly the most popular explanation for the cosmic accelerated expansion. Though the cosmological constant $\Lambda$ is the simplest model of dark energy, it can explain almost all the current observations well. So, the cosmological constant model (usually dubbed the $\Lambda$CDM model) is always viewed as the primary candidate of the cosmological standard model. However, the cosmological constant $\Lambda$ has always been facing significant challenges, such as the ``coincidence problem"~\cite{Peebles:2002gy}, which states that dark matter and dark energy just have the same order of magnitude today, while the order difference between them can be up to $10^{30}$ in the early universe. 

The ``interacting dark energy'' (IDE) scenario, in which it is considered that there is some direct, non-gravitational coupling between dark energy and dark matter, is proposed and studied widely \cite{He:2008tn,He:2009pd,Guo:2007zk,Boehmer:2008av,He:2009mz,Xia:2009zzb,Wei:2010cs,He:2010im,Li:2011ga,Li:2013bya,Murgia:2016ccp,Pourtsidou:2016ico,Costa:2016tpb,Xia:2016vnp,vandeBruck:2016hpz,Amendola:1999er,Comelli:2003cv,Zhang:2005rg,Cai:2004dk,Zimdahl:2005bk,Wang:2006qw,Amendola:2001rc,Bertolami:2007zm,Koyama:2009gd,Zhang:2004gc,Zhang:2007uh,Zhang:2009qa,Li:2010ak,Fu:2011ab,Zhang:2013lea,Geng:2015ara,Yin:2015pqa,Feng:2016djj,Wang:2016lxa,Kumar:2017dnp,Sola:2016jky,Sola:2016ecz,Sola:2016zeg,Sola:2017jbl,Li:2009zs,Zhang:2012uu}. It has been shown that the coincidence problem can be well alleviated in the IDE scenario \cite{He:2008tn,He:2009pd,Comelli:2003cv,Zhang:2005rg,Cai:2004dk}. But the more important question is how to detect such a direct coupling between dark energy and dark matter through the cosmological observations. This requires us to precisely know how the interaction affects the evolution of the universe, including the expansion history and the growth of structure of the universe. The impacts on the cosmic microwave background (CMB)~\cite{He:2009pd,Pourtsidou:2016ico} and large-scale structure formation~\cite{He:2009mz,Pourtsidou:2016ico,Amendola:2001rc,Bertolami:2007zm,Koyama:2009gd,Li:2013bya} in the IDE scenario have been investigated in detail.

The fact that neutrinos have masses leads to the phenomenon of neutrino oscillations. But, the absolute masses of neutrinos could not be measured by the neutrino oscillation experiments. Nevertheless, the neutrino mass (the total mass of neutrinos is denoted as $\sum m_\nu$) can leave imprints in the evolution of the universe, i.e., it has some subtle influences on the CMB anisotropies and matter clustering as well as the cosmological distances, and thus the absolute mass of neutrinos $\sum m_\nu$ can be probed by the cosmological observations. The current observations cannot tightly constrain the neutrino mass, but can only give an upper limit for $\sum m_\nu$ \cite{Hu:1997mj,Reid:2009nq,Thomas:2009ae,Carbone:2010ik,Audren:2012vy,Riemer-Sorensen:2013jsa,Font-Ribera:2013rwa,Palanque-Delabrouille:2014jca,Geng:2014yoa,Zhang:2015rha,Geng:2015haa,Chen:2015oga,Cuesta:2015iho,Chen:2016eyp,Moresco:2016nqq,Lu:2016hsd,Kumar:2016zpg,Ade:2015xua,Zhang:2014dxk,Zhang:2014nta,Zhang:2014ifa,Zhang:2014lfa,Li:2015poa,Zhang:2015uhk,Wang:2016tsz,Zhao:2016ecj,Xu:2016ddc,Li:2012vn,Wang:2012uf}. Recently, the Planck space mission released the new data of CMB anisotropies, which precisely constrain a lot of cosmological parameters and also give tight upper limits on the total neutrino mass \cite{Ade:2015xua}: $\sum m_{\nu}<0.72$ eV from Planck TT+lowP, $\sum m_{\nu}<0.21$ eV from Planck TT+lowP+BAO, $\sum m_{\nu}<0.49$ eV from Planck TT, TE, EE+lowP, and $\sum m_{\nu}<0.17$ eV from Planck TT, TE, EE+lowP+BAO, in the $\Lambda$CDM model; note that here ``lowP'' denotes the Planck low-$\ell$ polarization data. The issue concerning how the neutrino mass limit is influenced by the nature of dark energy has also been discussed in depth \cite{Zhang:2015uhk,Wang:2016tsz,Zhao:2016ecj}. In this paper, we wish to discuss how the interaction between dark energy and dark matter affects the cosmological weighing of neutrinos.

To extend the base $\Lambda$CDM model in the aspect of dark energy, one can consider the equation-of-state parameter of dark energy $w$, and further consider the coupling parameter between dark energy and dark matter $\beta$---this model can be called the I$w$CDM model. But, in this paper, we confine our discussions in the model of vacuum energy interacting with cold dark matter, abbreviated as the I$\Lambda$CDM model. The reasons include: (i) we should involve less parameters in the cosmological model, and (ii) we wish to see the pure impacts from the coupling. Therefore, in this study we only consider the vacuum energy with $w=-1$. Nevertheless, since the vacuum energy interacts with the cold dark matter, it is not a pure background any more; it must have perturbations, as the same as other energy components in the universe.\footnote{In the scenario of vacuum energy interacting with cold dark matter, the vacuum energy density does not maintain as a constant in space and time any more; actually, it is now a dynamical quantity [see Eq.~(\ref{2.1})]. In this case, the vacuum energy density is not equivalent to the cosmological constant $\Lambda$. That is to say, though $w=-1$ still holds for the vacuum energy, its density is no longer a constant that can characterize the spacetime background. Now that the vacuum energy density is a dynamical quantity in this scenario, it must have perturbations as the response to the metric fluctuations. In this paper, for convenience, such a scenario is still denoted as ``I$\Lambda$CDM'', although here the vacuum energy is actually not a $\Lambda$; one should keep in mind that this is only an abbreviation.} 

In order to treat the cosmological perturbation evolution in the I$\Lambda$CDM model, we employ the parameterized post-Friedmann (PPF) scheme for the IDE scenario \cite{Li:2014eha,Li:2014cee,Li:2015vla,Zhang:2017ize} to do the calculations. In this paper, we consider two cases of the energy transfer rate, i.e., $Q = \beta H \rho_{\rm c}$ and $Q = \beta H \rho_{\rm de}$, where $H$ is the Hubble expansion rate of the universe, and $ \rho_{\rm c}$ and $\rho_{\rm de}$ (in the case of I$\Lambda$CDM,  $\rho_{\rm de}$ is replaced with  $\rho_{\Lambda}$) denote the energy densities of cold dark matter and dark energy, respectively. We will discuss the observational constraints on the coupling $\beta$ and its impacts on the cosmological weighing of neutrinos. The base $\Lambda$CDM model acts as the reference model in this study.

The structure of the paper is organized as follows. In Sec.~\ref{model}, we describe in detail the two I$\Lambda$CDM models with the energy transfer rate $Q = \beta H \rho_{\rm c} $ and $Q = \beta H \rho_{\Lambda}$, respectively. The PPF approach is used to eliminate the large-scale instability in the IDE scenario. In Sec.~\ref{data}, we introduce the cosmological observational data used in this paper. In Sec.~\ref{results}, the constraint results are given for the $\Lambda$CDM model and the I$\Lambda$CDM models from different cosmological observations. When $\sum m_\nu$ is not treated as a free parameter, we set $\sum m_{\nu}=0.06$ eV;  when $\sum m_{\nu}$ acts as a free parameter, we set a flat prior $[0,3.0]$ eV for it. Conclusion is given in Sec.~\ref{Conclusion}.

\section{The scenario of vacuum energy interacting with cold dark matter under the PPF framework}\label{model}

The $\Lambda$CDM model can explain almost all the current observational data well. In particular, the latest Planck CMB data, combined with other astrophysical observations, strongly favor a base $\Lambda$CDM cosmology with only six parameters \cite{Ade:2015xua}. However, it is hard to believe that the evolution of the whole universe can be entirely described by only six parameters. It is believed that the 6-parameter base $\Lambda$CDM model is favored just because the current observations are not accurate enough. Thus, the base $\Lambda$CDM model would be extended, in the aspects of, e.g., dark energy, dark matter, fifth force, reionization, neutrino mass, primordial gravitational waves, and so on, and the future highly accurate observational data would offer evidence for some of these. When wishing to extend the base $\Lambda$CDM by adding one more parameter, the options include considering the addition of $w$ or $\beta$, corresponding to the $w$CDM model and the I$\Lambda$CDM model, respectively. The issue of weighing neutrinos in the $w$CDM cosmology (and other dark energy models) has been investigated in Refs. \cite{Zhang:2015uhk,Wang:2016tsz,Zhao:2016ecj}. Thus, in this work, we consider the latter, i.e., weighing neutrinos in the I$\Lambda$CDM cosmology. As mentioned above, we consider two cases of interaction form (energy transfer rate), $Q = \beta H \rho_{\rm c}$ and $Q = \beta H \rho_{\Lambda}$, in this paper.

For the I$\Lambda$CDM models, the energy conservation equations for vacuum energy (with density $\rho_{\Lambda}$) and cold dark matter (with density$\rho_{\rm c}$) are given by
\begin{equation}\label{2.1}
  {\rho}'_{\Lambda} = aQ_{\Lambda} ,
\end{equation}
\begin{equation}\label{2.2}
  {\rho}'_{\rm c} = -3{\cal H}\rho_{\rm c}+aQ_{\rm c} ,
\end{equation}
where $Q_{\Lambda}=-Q_{\rm c}=Q$, the prime is the derivative with respect to the conformal time $\eta$, and ${\cal H}$ is the conformal Hubble expansion rate, ${\cal H}= {a}'/a$ ($a$ is the scale factor of the universe). Once the form of $Q$ is specified, one can then obtain the background evolutions of vacuum energy and dark matter. From equations (\ref{2.1}) and (\ref{2.2}), we see that $\beta>0$ denotes cold dark matter decaying into vacuum energy, and $\beta<0$ denotes vacuum energy decaying into cold dark matter. When $\beta=0$, the I$\Lambda$CDM model reduces to the standard $\Lambda$CDM model.

Note here that when we consider the situation that vacuum energy interacts with cold dark matter, then the vacuum energy is some dynamical entity, and thus it is not a pure background any more. We must consider the perturbations of the vacuum energy in this case. The covariant conservation law is given by $\nabla_{\nu}T^{\mu \nu}_{I}=Q^{\mu}_{I}$ and $\sum_I Q^\mu_I=0$ ($I= \rm c, \Lambda$), in which $T^{\mu \nu}_{I}$ and $Q^{\mu}_{I}$ denote the energy-momentum tensor and the energy-momentum transfer vector, respectively. $Q^{\mu}_{I}$ can be split into two parts as follows,
\begin{equation}\label{2.3}
  Q^{I}_{\mu} = a(-Q_{I}(1+AY)-\delta Q_{I}Y,[f_{I}+Q_{I}(v-B)]Y_{i}),
\end{equation}
where $\delta Q_{I}$ and $f_{I}$ are the energy transfer perturbations and the momentum transfer potential of $I$ fluid. $A$ and $B$ are used to describe the scalar metric perturbations. $Y$ is the eigenfunctions of the Laplace operator, $\nabla^{2}Y= -k^{2}Y$ ($k$ is the comoving wave number), and $Y_{i}=(-k)\nabla_{i}Y$ is its covariant derivative. $v$ is the velocity of the total fluid. In order not to consider momentum transfer sector in the rest frame of cold dark matter, we take $Q^{\mu}_{\Lambda}=-Q^{\mu}_{\rm c}=Qu^{\mu}_{\rm c}$, in which $u^{\mu}_{\rm c}$ is the four-velocity of cold dark matter.

However, one should be warned about the perturbation instability problem appearing occasionally in some cases when calculating the dark energy perturbation evolution, in particular for the IDE scenario \cite{Valiviita:2008iv}. Also, in the present situation, we are treating the case of $w=-1$, which is difficult to consider the perturbation evolution according to the traditional linear perturbation theory of hydromechanics under general relativity. Therefore, we shall employ the PPF approach for the IDE scenario \cite{Li:2014eha,Li:2014cee} to calculate the perturbation evolution in the present study. This PPF scheme generalizes the previous PPF framework of uncoupled dark energy \cite{Hu:2008zd,Fang:2008sn} and can be used to cure the perturbation instability of dark energy in both the coupled and uncoupled cases. Using the PPF approach, without assuming any specific ranges of $w$ and $\beta$, the whole parameter space of the IDE scenario could be explored by the observational data. 

The main idea of the PPF description is to establish an effective theory to treat the dark energy perturbation totally based on the basic facts of dark energy---since the nature of dark energy is obscure, one should not consider the dark energy pressure perturbation related to the rest-frame sound speed imposed by hand. In the PPF framework, the perturbations of dark energy are decribed by some parameterizations for the scales of far beyond the horizon and deep inside the horizon, respectively, and then a well-behaved continuous function is found to link these two limits. It has been shown that the PPF method can give stable cosmological perturbations in the IDE scenario for any cases (see Refs.~\cite{Li:2014eha,Li:2014cee}). In the following, we give a brief description of the PPF framework for the IDE scenario. For more details, we refer the reader to Refs.~\cite{Li:2014eha,Li:2014cee}.

The conservation equations for the $I$ fluid in the IDE scenario can be explicitly given by
\begin{equation}
 {\delta\rho_I'}
	+  3\mathcal{H}({\delta \rho_I}+ {\delta p_I})+(\rho_I+p_I)(k{v}_I + 3 H_L')=a(\delta Q_I+AQ_I),\label{eqn:conservation1}
\end{equation}	
\begin{equation}	
[(\rho_I + p_I)({{v_I}-{B}})]'+4\mathcal{H}(\rho_I + p_I)({{v_I}-{B}}) -k{ \delta p_I }+ {2 \over 3}k\ck p_I {\Pi_I} - k(\rho_I+ p_I) {A}=a[Q_I(v-B)+f_I],\label{eqn:conservation2}
\end{equation}
where $\delta\rho_I$ is energy density perturbation, $v_I$ is velocity perturbation, $\delta p_I$ is isotropic pressure perturbation, $\Pi_I$ is anisotropic stress perturbation, and $c_K = 1-3K/k^2$ with $K$ the spatial curvature.

The discussion of the perturbation evolutions of cold dark matter and dark energy is made in the comoving gauge, i.e., $B=v_T$ and $H_T=0$, where $v_T$ denotes the velocity perturbation of total matters except dark energy. To avoid confusion, we use the new symbols in this discussion---$\zeta\equiv H_L$, $\xi\equiv A$, $\rho\Delta\equiv\delta\rho$, $\Delta p\equiv\delta p$, $V\equiv v$, and $\Delta Q_I\equiv\delta Q_I$---to denote the corresponding quantities of the comoving gauge except for the two gauge independent quantities $\Pi$ and $f_I$. For the cold dark matter, we have $\Delta p_c=\Pi_c=0$, and thus the evolutions of the remaining two quantities $\rho_c\Delta_c$ and $V_c$ are totally determined by Eqs.~(\ref{eqn:conservation1}) and (\ref{eqn:conservation2}). Note that $\Delta Q_{I}$ and $f_{I}$ can be easily derived in a specific IDE model. For the dark energy, we need an extra condition for $\Delta p_{\rm de}$ besides $\Pi_{\rm de}=0$ and Eqs.~(\ref{eqn:conservation1}) and (\ref{eqn:conservation2}) to complete the dark energy perturbation system. If we treat dark energy as a nonadiabatic fluid and calculate $\Delta p_{\rm de}$ in terms of the adiabatic sound speed and the rest frame sound speed (see, e.g., Ref.~\cite{Valiviita:2008iv}), then the large-scale instability in the IDE scenario will occur. Thus, we need to treat the dark energy perturbations within the generalized PPF framework established in Refs.~\cite{Li:2014eha,Li:2014cee}. 

The key point is to establish a direct relationship between $V_{\rm de} - V_T$ and $V_T$ on the large scales instead of directly defining a rest-frame sound speed for dark energy and calculating $\Delta p_{\rm de}$ based on it. This relationship can be parametrized by a function $f_\zeta(a)$ as \cite{Hu:2008zd,Fang:2008sn}
\begin{equation}
\lim_{k_H \ll 1}
 {4\pi G a^2\over \mathcal{H}^2} (\rho_{\rm de} + p_{\rm de}) {V_{\rm de} - V_T \over k_H}
= - {1 \over 3} \ck  f_\zeta(a) k_H V_T,\label{eq:DEcondition}
\end{equation}
where $k_H=k/\mathcal{H}$. This condition combined with the Einstein equations gives the equation of motion for the curvature perturbation $\zeta$ on the large scales,
\begin{align}
\lim_{k_H \ll 1} \zeta'  = \mathcal{H}\xi - {K \over k} V_T +{1 \over 3} \ck  f_\zeta(a) k V_T.
\label{eqn:zetaprimesh}
\end{align}
On the small scales, the evolution of the curvature perturbation is described by the Poisson equation, $\Phi=4\pi G a^2\Delta_T \rho_T/( k^2\ck)$, with $\Phi=\zeta+V_T/k_H$. The evolutions of the curvature perturbation at $k_H\gg1$ and $k_H\ll1$ can be related by introducing a dynamical function $\Gamma$ to the Poisson equation, such that
\begin{equation}
\Phi+\Gamma = {4\pi Ga^2
\over  k^2\ck} \Delta_T \rho_T
\label{eqn:modpoiss}
\end{equation}
on all scales. 

Compared with the small-scale Poisson equation, Eq.~(\ref{eqn:modpoiss}) gives $\Gamma\rightarrow0$ at $k_H\gg1$. On the other hand, with the help of the Einstein equations and the conservation equations as well as the derivative of Eq.~(\ref{eqn:modpoiss}), Eq.~(\ref{eqn:zetaprimesh}) gives the equation of motion for $\Gamma$ on the large scales,
\begin{equation}\label{eq:gammadot}
\lim_{k_H \ll 1} \Gamma'  = S -\mathcal{H}\Gamma,
\end{equation}
with
\begin{align}
S&={4\pi Ga^2
\over k^2 } \Big\{[(\rho_{\rm de}+p_{\rm de})-f_{\zeta}(\rho_T+p_T)]kV_T \nonumber\\
&\quad+{3a\over k_Hc_K}[Q_c(V-V_T)+f_c]+\frac{a}{c_K}(\Delta Q_c+\xi Q_c)\Big\},\nonumber
\end{align}
where $\xi$ can be obtained from Eq.~(\ref{eqn:conservation2}),
\begin{equation}
\xi =  -{\Delta p_T - {2\over 3}\tppi+{a\over k}[Q_c(V-V_T)+f_c] \over \rho_T + p_T}.
\label{eqn:xieom}
\end{equation}

Using a transition scale parameter $c_\Gamma$, we can take the equation of motion for $\Gamma$ on all scales to be \cite{Hu:2008zd,Fang:2008sn}
\begin{equation}
(1 + c_\Gamma^2 k_H^2) [\Gamma' +\mathcal{H} \Gamma + c_\Gamma^2 k_H^2 \mathcal{H}\Gamma] = S.
\label{eqn:gammaeom}
\end{equation}

From the above equations, we can find that all the perturbation quantities relevant to the equation of motion of $\Gamma$ are those of matters other than dark energy. Therefore, we can solve the differential equation (\ref{eqn:gammaeom}) without any knowledge of the dark energy perturbations. Once the evolution of $\Gamma$ is obtained, we can immediately obtain energy density and velocity perturbations of dark energy,
\begin{align}
&\rho_{\rm de}\Delta_{\rm de} =- 3(\rho_{\rm de}+p_{\rm de}) {V_{\rm de}-V_{T}\over k_{H} }-{k^{2}\ck \over 4\pi G a^{2}} \Gamma,\\ \label{eqn:ppffluid}
& V_{\rm de}-V_{T} ={-k \over 4\pi Ga^2 (\rho_{\rm de} + p_{\rm de}) F} \nonumber \\
&\quad\quad\quad\times\left[ S - \Gamma' - \mathcal{H}\Gamma + f_{\zeta}{4\pi Ga^2 (\rho_{T}+p_{T}) \over k}V_{T}
\right],
\end{align}
with $F = 1 +  12 \pi G a^2 (\rho_T + p_T)/( k^2 \ck)$.

In this paper, we use this generalized PPF approach in the I$\Lambda$CDM scenario. We only need to take $w=-1$ and replace the subscript ``de'' with ``$\Lambda$'' in this framework.

\section{Data and method}\label{data}

\begin{table*}[htbp]\tiny
\begin{center}
\caption{Priors on the free parameters for the I$\Lambda$CDM models in a flat universe. }
\label{table1}
\small
\setlength\tabcolsep{9.8pt}
\renewcommand{\arraystretch}{1.2}
\begin{tabular}{cccccccccc}
\\
\hline\hline
Paramerer  &&&&&& Prior   \\ \hline

$\Omega_{\rm b}h^{2}$   &&&&&& $[0.005,0.100]$    \\
$\Omega_{\rm c}h^{2}$   &&&&&&  $[0.001,0.990]$    \\
$100\theta_{\rm MC}$  &&&&&&  $[0.5,10.0]$    \\
$\tau$    &&&&&&$[0.01,0.80]$    \\
$\ln(10^{10}A_{\rm s})$   &&&&&& $[2,4]$    \\
$n_{\rm s}$   &&&&&& $[0.8,1.2]$   \\
$\beta$ ($Q=\beta H \rho_{\rm c}$)   &&&&&& $[-0.015,0.050]$     \\
$\beta$ ($Q=\beta H \rho_{\Lambda}$)   &&&&&& $[-1.0,0.8]$     \\
\hline
$\sum m_{\nu}$  &&&&&&    $[0,3.0]$ eV   \\
\hline\hline
\end{tabular}
\end{center}
\end{table*}

In this paper, we use the latest cosmological observational data to constrain the neutrino mass $\sum m_\nu$, the coupling parameter $\beta$, together with other cosmological parameters. In what follows, we shall first briefly describe the observational data.

\textbf{Planck}: The neutrino mass has some subtle influences on the evolution of the universe, and so it will leave some imprints in the CMB spectra. The Planck space mission has measured the CMB anisotropies with unprecedented accuracy. We use the full Planck data released in 2015~\cite{Ade:2015xua}, including the power spectrum of temperature (TT), the power spectrum of polarization E-mode (EE), the cross-correlation power spectrum of temperature and E-mode (TE), and the low-$\ell$ ($\ell \leq 30$) temperature-polarization spectrum (lowP). This data set is abbreviated as ``Planck" in this paper.

\textbf{BSH}: The baryon acoustic oscillation (BAO) and type Ia supernova (SNIa) data are in excellent agreement with the latest Planck data, and thus in this paper we use these observations to help improve constraints on cosmological parameters. We use four BAO points: the CMASS and LOWZ samples from the latest Data Release 12 (DR12) of the Baryon Oscillation Spectroscopic Survey (BOSS) at $z_{\rm eff}=0.57$ and $z_{\rm eff}=0.32$~\cite{Gil-Marin:2016wya}, respectively, the 6dF Galaxy Survey (6dFGS) measurement at $z_{\emph{\emph{eff}}}=0.106$~\cite{Beutler:2011hx}, and the Main Galaxy Sample of Data Release 7 of Sloan Digital Sky Survey (SDSS-MGS) at $z_{\emph{\emph{eff}}}=0.15$~\cite{Ross:2014qpa}. For the SNIa data we use the ``Joint Light-curve Analysis" (JLA) compilation, which is comprised of SNLS and SDSS together with several samples of low redshift light-curve analysis.  We also combine the latest local measurement of the Hubble constant, $H_{0}=73.00\pm1.75$ km s$^{-1}$ Mpc$^{-1}$, whose uncertainty has been reduced from 3.3\% to 2.4\% by using the WFC3 in the HST~\cite{Riess:2016jrr}. In addition, the BAO data from the DR11~\cite{Anderson:2013zyy} and the Hubble constant measurement $H_{0}=70.6\pm3.3$ km s$^{-1}$ Mpc$^{-1}$~\cite{Efstathiou:2013via} are also used to make a comparison with those from the DR12 and $H_{0}=73.00\pm1.75$ km s$^{-1}$ Mpc$^{-1}$. Note also that BAO stands for the BAO data involving the DR12 and $H_{0}$ refers to the measurement by Riess et.al in 2016, without other statement in this paper. For simplicity, ``BSH" is used to denote the combination of BAO, SNIa, and Hubble constant $H_{0}$.

\textbf{LSS}: Furthermore, we consider the large-scale structure (LSS) observations to constrain the neutrino mass for the $Q = \beta H \rho_{\Lambda}$ model. Due to the direct correlation between peculiar velocities of galaxies in the linear perturbation theory and matter clustering, the redshift-space distortions (RSD) observation could give the measurement of $f\sigma_{8}$, which is the product of the linear structure growth rate parameter $f$ and the amplitude of matter density fluctuation $\sigma_{8}$~\cite{Peacock:2001gs,Guzzo:2008ac}. References~\cite{Li:2014cee,Yang:2014gza,Li:2015vla} have shown that the RSD data can help tightly constrain the coupling strength for the $Q=\beta H  \rho_{\Lambda}$ model. So, we employ two latest RSD measurements to constrain the $Q=\beta H \rho_{\Lambda}$ model in this paper. They are from the CMASS sample with an effective redshift of $z_{\rm eff}=0.57$ and the LOWZ sample with an effective redshift of $z_{\rm eff}=0.32$~\cite{Gil-Marin:2016wya}, respectively. Reference~\cite{Gil-Marin:2016wya} demonstrates general agreement between the DR11~\cite{Anderson:2013zyy} and DR12 measurements. In addition, we also consider the weak lensing (WL) data from CFHTLenS~\cite{Heymans:2013fya} in this work. We use the CFHTLenS tomographic blue galaxy sample \cite{Heymans:2013fya} whose shear correlation functions are estimated in six redshift bins, each with an angular range $1.7'<\theta<37.9'$. Following the Planck collaboration \cite{Ade:2015rim}, we choose to adopt the ``ultra-conservative'' cuts (i.e., remove $\xi^-$ entirely from each dataset and exclude $\theta<17'$ for $\xi^+$) to remain the robustness for the tomographic data to nonlinear modelling, baryonic feedback, and intrinsic alignment marginalization. In this paper, we use ``LSS" to stand for the combination of RSD and WL data.

We mainly consider two combinations of observational data in our analysis: Planck+BSH and Planck+BSH+LSS. We do not use the LSS data for the $\Lambda$CDM model and the $Q=\beta H \rho_{\rm c}$ model because there is a strong tension on the amplitude of matter fluctuation spectrum between Planck and LSS. However, the RSD data can provide a tighter constraint for the $Q=\beta H \rho_{\Lambda}$ model as mentioned in Refs. \cite{Li:2014cee,Yang:2014gza,Li:2015vla}. Thus we employ the LSS data in this situation.

In our work, we have a great interest in how the coupling between dark energy and dark matter affects the cosmological constraints on the neutrino mass. We assume a degenerate mass model ($m_{1}=m_{2}=m_{3}$) regardless of the mass splitting for the three neutrino mass eigenstates in our calculations, as the same as the Planck collaboration. Here the neutrino mass is set as a free parameter, whose prior is $[0,3.0]$ eV in a flat universe. When we do not consider the influence from neutrinos, we uniformly take two massless and one massive neutrino species with total mass $\sum m_{\nu} = 0.06$ eV.

We give the flat prior of every free parameter for the two I$\Lambda$CDM models, as shown in Table \ref{table1}. Here $\Omega_{\rm b}h^{2}$ and $\Omega_{\rm c}h^{2}$ stand for the physical baryon and cold dark matter densities, respectively. $\theta_{\rm MC}$ is the ratio of the sound horizon $r_{\rm s}$ to the angular diameter distance $D_{\rm A}$ at last-scattering time. $\tau$ denotes the optical depth to reionization. And $n_{\rm s}$ and $A_{\rm s}$ indicate the spectral index and the amplitude of the primordial power spectrum of scalar perturbations, respectively. For the coupling constant $\beta$, we set the prior of $[-0.015,0.050]$ and $[-1.0,0.8]$ for the $Q=\beta H \rho_{\rm c}$  and $Q=\beta H \rho_{\Lambda}$ models, respectively. Such priors are reasonable to probe the whole parameter space of $\beta$. We have empirically found that the parameter space of $\beta$ is not symmetric; of course, one can set a larger symmetric range for it, but the results will be the same. To correctly solve the background and perturbation equations for the I$\Lambda$CDM models, we make some changes for the PPF code \cite{Hu:2008zd,Fang:2008sn,Li:2014eha,Li:2014cee,Li:2015vla} and the Boltzmann code CAMB \cite{Lewis:1999bs}. We also use the Markov-chain Monte Carlo package CosmoMC \cite{Lewis:2002ah} to probe the whole parameter space.

\section{Results}\label{results}

In this section, we will first give the updated upper bound on the neutrino mass for the $\Lambda$CDM model from Planck+BSH. 
Here the ``updated'' upper bound means that we compare with the previous result in Ref.~\cite{Zhao:2016ecj}. In Ref.~\cite{Zhao:2016ecj}, the neutrino mass is also constrained in $\Lambda$CDM cosmology using Planck+BSH, but in this case the BAO BOSS DR11 data~\cite{Anderson:2013zyy} and the Hubble constant prior $H_{0}=70.6\pm3.3$ km s$^{-1}$ Mpc$^{-1}$~\cite{Efstathiou:2013via} are used, which gives the constraint result $\sum m_\nu<0.15$ eV. 

Then, we give the constraint results on the coupling constant $\beta$ and the neutrino mass $\sum m_{\nu}$, and analyze how the coupling constant $\beta$ affects the cosmological constraints on the neutrino mass in the $Q=\beta H  \rho_{\rm c}$ and $Q=\beta H  \rho_{\Lambda}$ models.

\subsection{Constraints on the neutrino mass in the base $\Lambda$CDM model}

\begin{table*}[ht!]\tiny
\caption{Fit results for the $\Lambda$CDM and $\Lambda$CDM+$\sum m_{\nu}$ models by using Planck+BSH.}
\label{table2plus}
\small
\setlength\tabcolsep{2.8pt}
\renewcommand{\arraystretch}{1.2}
\centering
\begin{tabular}{cccccccccccc}
\\
\hline\hline &\multicolumn{1}{c}{$\Lambda$CDM} &&&&&&&\multicolumn{1}{c}{$\Lambda$CDM+$\sum m_{\nu}$}
\\ \hline

$\sum m_{\nu}$               &...&&&&&&
                   & $<0.11$\\

$\Omega_{\rm m}$       & $0 .3033^{+0.0058}_{-0.0056}$&&&&&&
                   & $0.3020^{+0.0060}_{-0.0061}$
                   \\

$H_{0}$         & $68 .17^{+0.43}_{-0.44}$&&&&&&
                   & $68.31\pm0.47$
                   \\

$\sigma_{8}$           & $0 .832^{+0.013}_{-0.014}$&&&&&&
                  & $0.837\pm0.015$
                   \\
\hline
$\chi^{2}_{\rm min}$        & $13668.97$&&&&&&
                   &$13665.92$
\\
\hline\hline
\end{tabular}
\end{table*}

\begin{figure*}[ht!]
\begin{center}
\includegraphics[width=11cm]{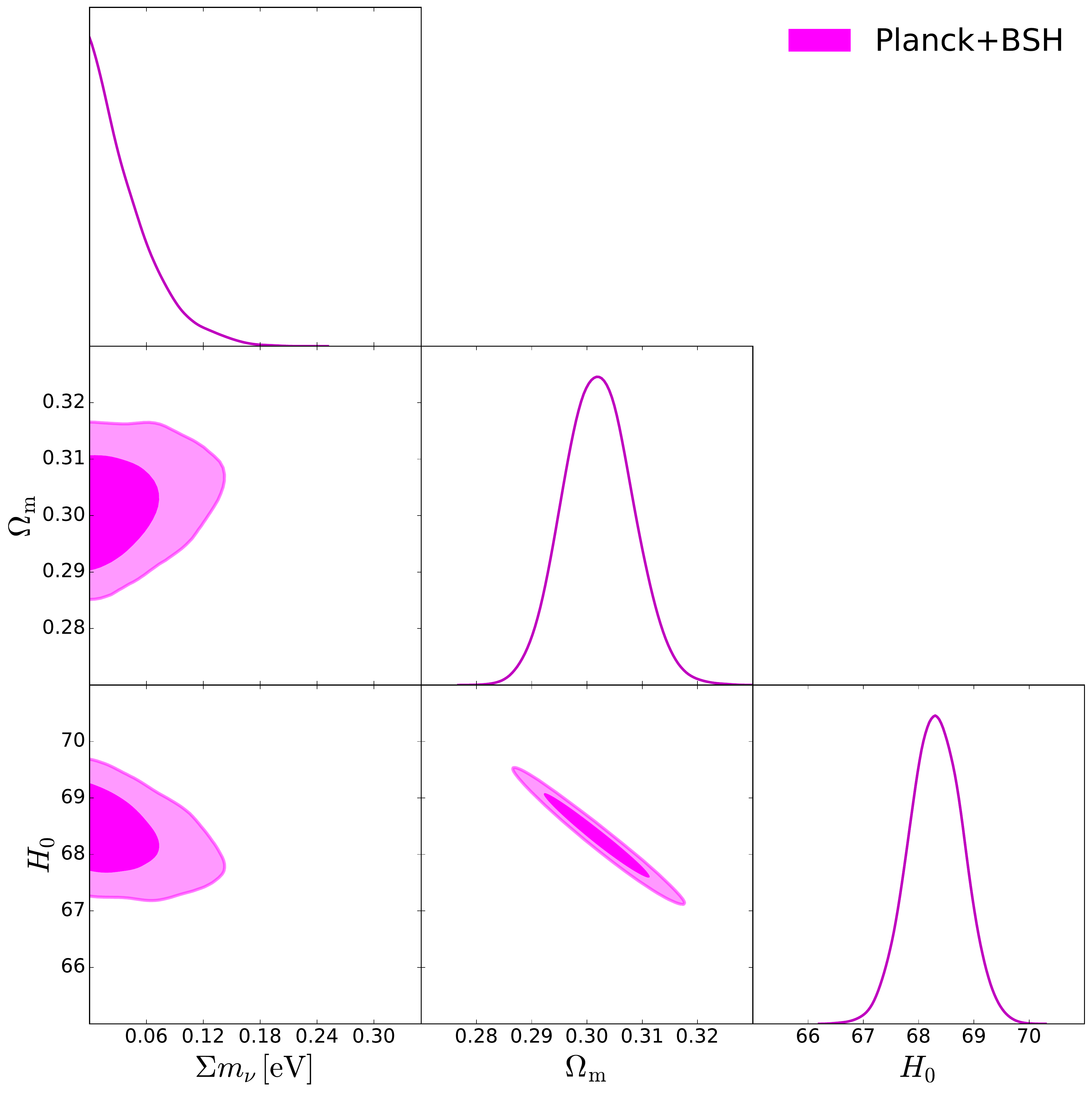}
\end{center}
\vspace{-8mm}
\caption{One-dimensional marginalized distributions and two-dimensional contours at $1\sigma$ and $2\sigma$ level for parameters $\sum m_{\nu}$, $\Omega_{\rm m}$, and $H_{0}$ of the $\Lambda$CDM+$\sum m_{\nu}$ model by using Planck+BSH.}
\label{fig1}
\end{figure*}
 
The $\Lambda$CDM model is taken as the baseline model in our study, and in this model the upper limit on the neutrino mass is given: $\sum m_{\nu}<0.11$ eV at $2\sigma$ level from Planck+BSH. We know that there is a lower limit of the neutrino mass for the inverted hierarchy ($m_{3}\ll m_{1}<m_{2}$) of neutrino mass spectrum---$\sum m_\nu>0.1$ eV \cite{olive}, and thus the small upper limit result, $\sum m_{\nu}<0.11$ eV, derived in this paper, is rather close to its lower cut-off value, which implies that the cosmological weighing of neutrinos is now close to the edge of diagnosing the mass hierarchies of neutrinos. Further more accurate data might be able to give clear evidence. 

Here we make a further analysis for this result. We give the constraint $\sum m_{\nu}<0.15$ eV from Planck+BAO, where BAO data is from the DR12. Compared to the result $\sum m_{\nu}<0.17$ eV~\cite{Ade:2015xua} from Planck+BAO, where BAO data is from the DR11, we find that the DR12 data can provide a 12\% reduction to the upper limit of neutrino mass, and thus combining with the DR12 can indeed lead to a smaller upper limit on the neutrino mass. Similar results and analyses can be found in Refs. \cite{Huang:2015wrx,Zhang:2015uhk,Wang:2016tsz,Zhao:2016ecj}.

In addition, the small upper limit $\sum m_{\nu}<0.11$ eV derived may be due to the fact that we use the latest Hubble constant measurement result $H_{0}=73.00\pm1.75$ km s$^{-1}$ Mpc$^{-1}$~\cite{Riess:2016jrr}. This new $H_0$ measurement is $3\sigma$ higher than the Planck fit result $H_{0}=67.27\pm0.66$ km s$^{-1}$ Mpc$^{-1}$~\cite{Ade:2015xua} based on the $\Lambda$CDM model. The strong tension may lead to a lower neutrino mass. To confirm the inference, we give the constraint $\sum m_{\nu}<0.13$ eV from Planck+BSH, where the local measurement value $H_{0}=70.6\pm3.3$ km s$^{-1}$ Mpc$^{-1}$ is used, which is only $1\sigma$ higher than $H_{0}=67.27\pm0.66$ km s$^{-1}$ Mpc$^{-1}$. The result shows that $H_{0}=73.00\pm1.75$ km s$^{-1}$ Mpc$^{-1}$ indeed allows a lower neutrino mass. Debates about the $H_{0}$ measurements have been discussed in Refs. \cite{Efstathiou:2013via,Riess:2011yx,Freedman:2012ny,Bernal:2016gxb} and it is necessary for us to do further research. However, we still employ $H_{0}=73.00\pm1.75$ km s$^{-1}$ Mpc$^{-1}$ due to its reduced uncertainty from 3.3\% to 2.4\% and tight constraints on other parameters in the following analysis.

 In Fig.~\ref{fig1}, we show the one-dimensional posterior distribution and two-dimensional contours at $1\sigma$ and $2\sigma$ levels for the parameters $\Omega_{\rm m}$, $\sum m_{\nu}$, and $H_{0}$ in the $\Lambda$CDM+$\sum m_{\nu}$ model by using Planck+BSH. From this figure, we clearly see that there is an anti-correlation between $\sum m_{\nu}$ and $H_{0}$. Therefore, the fact that $H_{0}=73.00\pm1.75$ km s$^{-1}$ Mpc$^{-1}$ is larger than $H_{0}=70.6\pm3.3$ km s$^{-1}$ Mpc$^{-1}$ would also lead to a lower neutrino mass. In addition, the data combination Planck+BSH gives the constraints: $\Omega_{\rm m}=0.3020^{+0.0060}_{-0.0061}$, $\emph{$H_{0}$}=68.31\pm0.47$ km s$^{-1}$ Mpc$^{-1}$, and $\sigma_{8}=0.837\pm0.015$ for the $\Lambda$CDM+$\sum m_{\nu}$ model.

\subsection{Constraints on the neutrino mass in the interacting vacuum energy models}

\begin{table*}[ht!]\tiny
\caption{Fit results for the I$\Lambda$CDM1 ($Q=\beta H\rho_{\rm c}$) and I$\Lambda$CDM1 ($Q=\beta H\rho_{\rm c}$)+$\sum m_{\nu}$ models by using Planck+BSH.}
\label{table2}
\small
\setlength\tabcolsep{2.8pt}
\renewcommand{\arraystretch}{1.2}
\centering
\begin{tabular}{cccccccccccc}
\\
\hline\hline &\multicolumn{1}{c}{I$\Lambda$CDM1} &&&&&&&\multicolumn{1}{c}{I$\Lambda$CDM1+$\sum m_{\nu}$}
\\ \hline

$\beta$                  &$0 .0021\pm0.0011$&&&&&&
                         &$0 .0021^{+0.0012}_{-0.0013}$\\

$\sum m_{\nu}$               &...&&&&&&
                   & $<0.20$\\

$\Omega_{\rm m}$       & $0.2943^{+0.0075}_{-0.0076}$&&&&&&
                   & $0.2945^{+0.0074}_{-0.0082}$
                   \\

$H_{0}$         & $68 .94^{+0.63}_{-0.62}$&&&&&&
                   & $68 .93\pm0.64$
                   \\

$\sigma_{8}$           & $0 .845\pm0.015$&&&&&&
                  & $0 .844^{+0.019}_{-0.017}$
                   \\
\hline
$\chi^{2}_{\rm min}$        & $13665.78$&&&&&&
                   &$13664.71$
\\
\hline\hline
\end{tabular}
\end{table*}

\begin{figure}[ht!]
\begin{center}
\includegraphics[width=7.5cm]{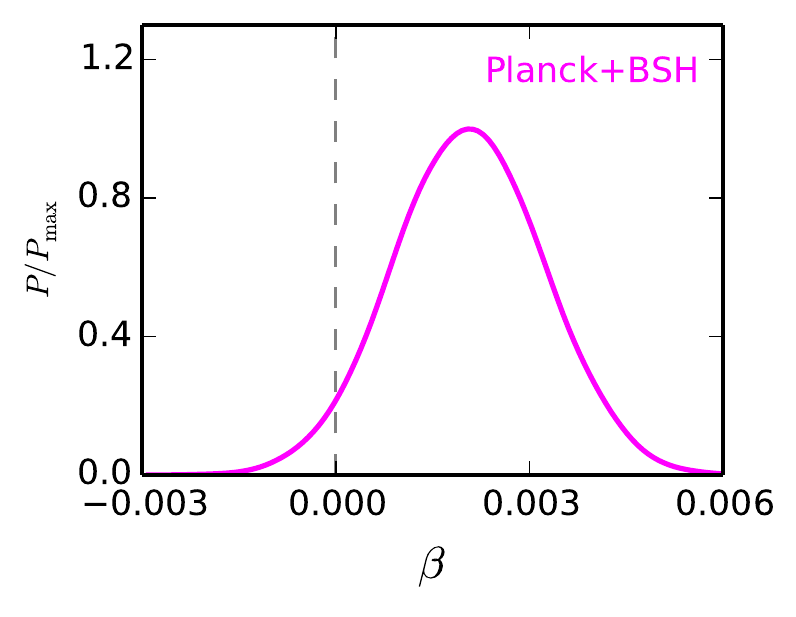}
\includegraphics[width=7.0cm]{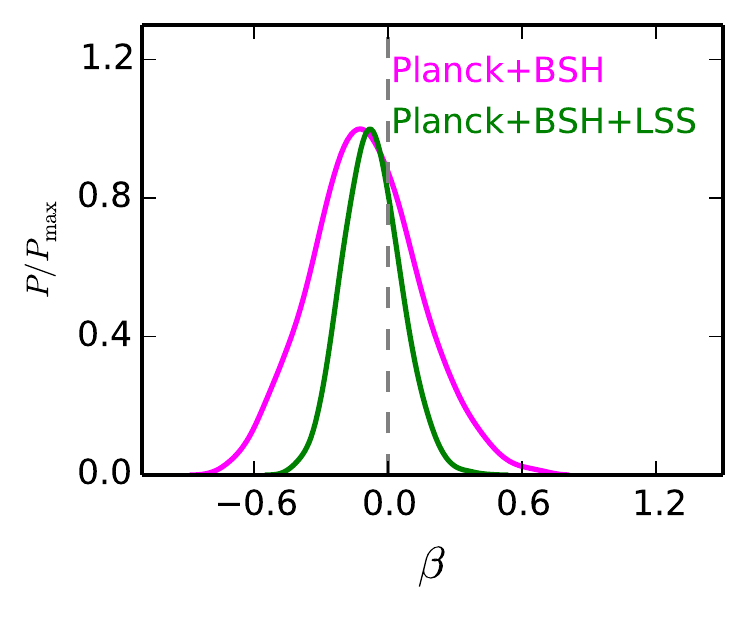}
\end{center}
\vspace{-8mm}
\caption{The one-dimensional posterior distributions for the coupling parameter $\beta$ in the $Q=\beta H  \rho_{\rm c}$ (\emph{left}) and $Q=\beta H  \rho_{\Lambda}$ (\emph{right}) models, with fixed neutrino mass $\sum m_\nu=0.06$ eV. }
\label{fig2}
\end{figure}

\begin{figure*}[ht!]
\begin{center}
\includegraphics[width=13cm]{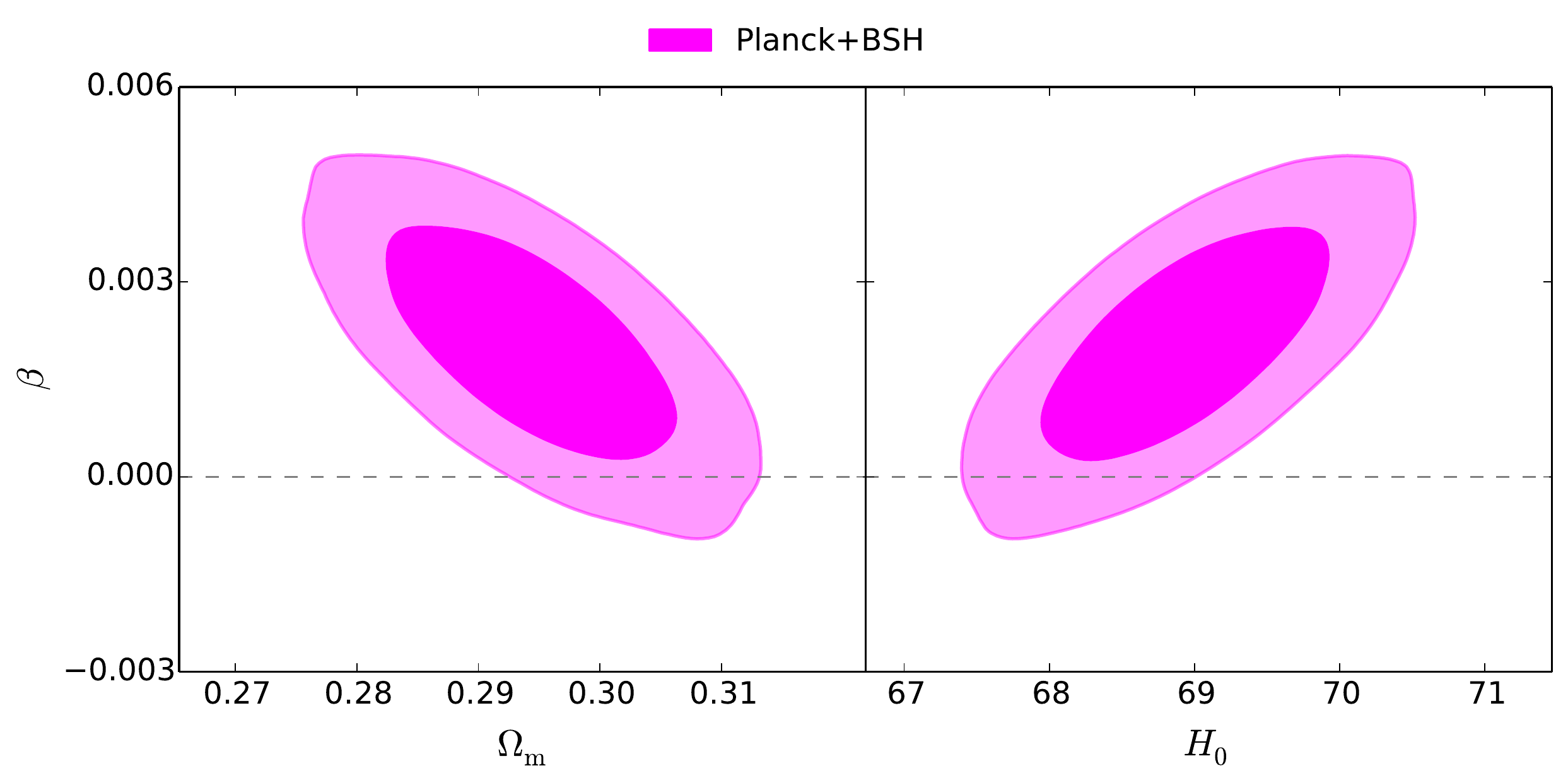}
\end{center}
\vspace{-8mm}
\caption{The two-dimensional marginalized contours ($1\sigma$ and $2\sigma$) in the $\beta$--$\Omega_{\rm m}$ and $\beta$--$H_0$ planes for the $Q=\beta H  \rho_{\rm c}$ model (with fixed neutrino mass $\sum m_\nu=0.06$ eV) by using Planck+BSH.}
\label{fig3}
\end{figure*}

\begin{figure*}[ht!]
\begin{center}
\includegraphics[width=11cm]{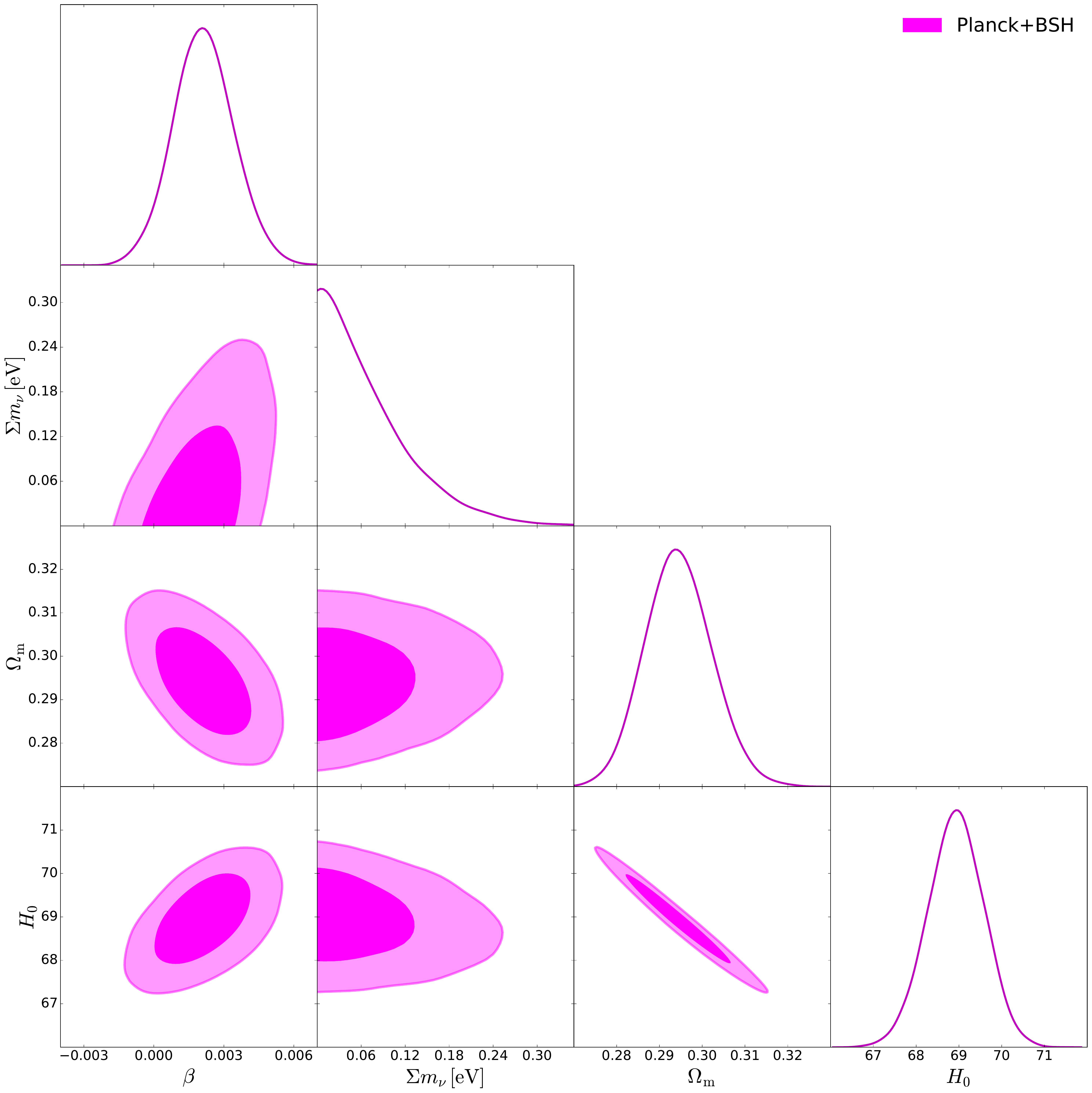}
\end{center}
\vspace{-8mm}
\caption{One-dimensional marginalized distributions and two-dimensional contours at $1\sigma$ and $2\sigma$ levels for parameters $\beta$, $\sum m_{\nu}$, $\Omega_{\rm m}$, and $H_{0}$ of the $Q=\beta H \rho_{\rm c}$ model in the presence of $\sum m_{\nu}$ by using Planck+BSH.}
\label{fig4}
\end{figure*}

Table \ref{table2} gives the detailed fitting results of the $Q=\beta H \rho_{\rm c}$ model for the cases without and with $\sum m_\nu$, from Planck+BSH. In this paper, all the fitting results are given at $1\sigma$ level, except for $\sum m_{\nu}$ (upper limit at the $2\sigma$ level). For simplicity, I$\Lambda$CDM1 and I$\Lambda$CDM2 are used to denote the $Q=\beta H \rho_{\rm c}$ and $Q=\beta H  \rho_{\Lambda}$ models, respectively, in all tables.  Obviously, the result shows that $\beta>0$ at more than 1$\sigma$ level ($\beta=0 .0021\pm0.0011$ and $\beta=0.0021^{+0.0012}_{-0.0013}$) no matter if we involve $\sum m_{\nu}$ or not in the $Q=\beta H \rho_{\rm c}$ model. Namely, the combination of Planck+BSH favors the case of cold dark matter decaying into vacuum energy. The results of $\beta>0$ (at more than 1$\sigma$) can be clearly seen from Figs.~\ref{fig2}--\ref{fig4}. In addition, for this model, the constraint result of neutrino mass is $\sum m_{\nu}<0.20$ eV. Figure \ref{fig4} shows a positive correlation between $\beta$ and $\sum m_{\nu}$. Thus, a larger $\beta$ will lead to a larger $\sum m_\nu$, which explains why the limit of $\sum m_\nu$ derived in this case is larger than that in $\Lambda$CDM.

\begin{table*}[ht!]\tiny
\caption{Fit results for the I$\Lambda$CDM2 ($Q=\beta H \rho_{\Lambda}$) and I$\Lambda$CDM2 ($Q=\beta H \rho_{\Lambda}$)+$\sum m_{\nu}$ models by using Planck+BSH and Planck+BSH+LSS.}
\label{table3}
\small
\setlength\tabcolsep{2.8pt}
\renewcommand{\arraystretch}{1.2}
\centering
\begin{tabular}{cccccccccccc}
\\
\hline\hline &\multicolumn{2}{c}{I$\Lambda$CDM2} &&\multicolumn{2}{c}{I$\Lambda$CDM2+$\sum m_{\nu}$} \\
           \cline{2-3}\cline{5-6}
Data  & Planck+BSH & Planck+BSH+LSS && Planck+BSH & Planck+BSH+LSS  \\ \hline

$\beta$                  &$-0.10\pm0.24$
                         &$-0.07^{+0.12}_{-0.14}$&
                         &$-0.04^{+0.20}_{-0.30}$
                         &$-0.08^{+0.13}_{-0.15}$\\

$\sum m_{\nu}$               &...
                               &...&
                   & $<0.10$
                   & $<0.14$\\

$\Omega_{\rm m}$       &$0 .328\pm0.055$
                   &$0 .32\pm0.029$&
                   &$0 .316^{+0.067}_{-0.049}$
                   &$0.322^{+0.034}_{-0.030}$
                   \\

$H_{0}$         &$68 .01^{+0.54}_{-0.50}$
                &$68 .14^{+0.47}_{-0.46}$&
                &$68 .19^{+0.55}_{-0.49}$
                &$68.22^{+0.56}_{-0.51}$
                   \\

$\sigma_{8}$           &$0 .816^{+0.038}_{-0.031}$
                       &$0 .808^{+0.019}_{-0.018}$&
                       &$0.827\pm0.035$
                       &$0 .809\pm0.020$
                   \\
\hline
$\chi^{2}_{\rm min}$        & $13668.65$
                   & $13701.15$&
                   &$13664.57$
                   &$13699.45$
\\
\hline\hline
\end{tabular}
\end{table*}

\begin{figure*}[ht!]
\begin{center}
\includegraphics[width=11cm]{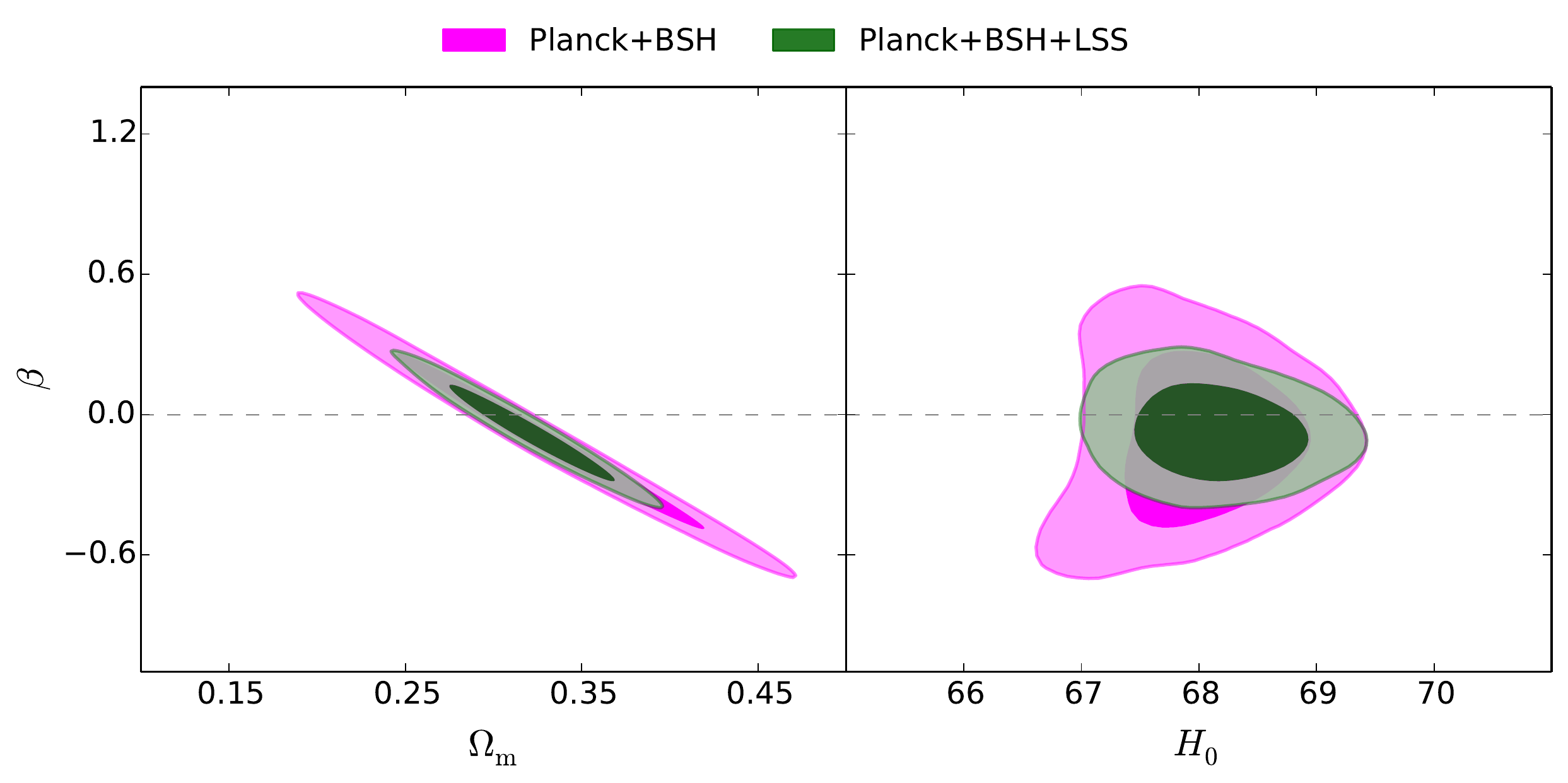}
\end{center}
\caption{The two-dimensional marginalized contours ($1\sigma$ and $2\sigma$) in the $\beta$--$\Omega_{\rm m}$ and $\beta$--$H_0$ planes for the $Q=\beta H  \rho_{\Lambda}$ model (with fixed neutrino mass $\sum m_\nu=0.06$ eV) by using Planck+BSH and Planck+BSH+LSS.}
\label{fig5}
\end{figure*}

\begin{figure*}[ht!]
\begin{center}
\includegraphics[width=13cm]{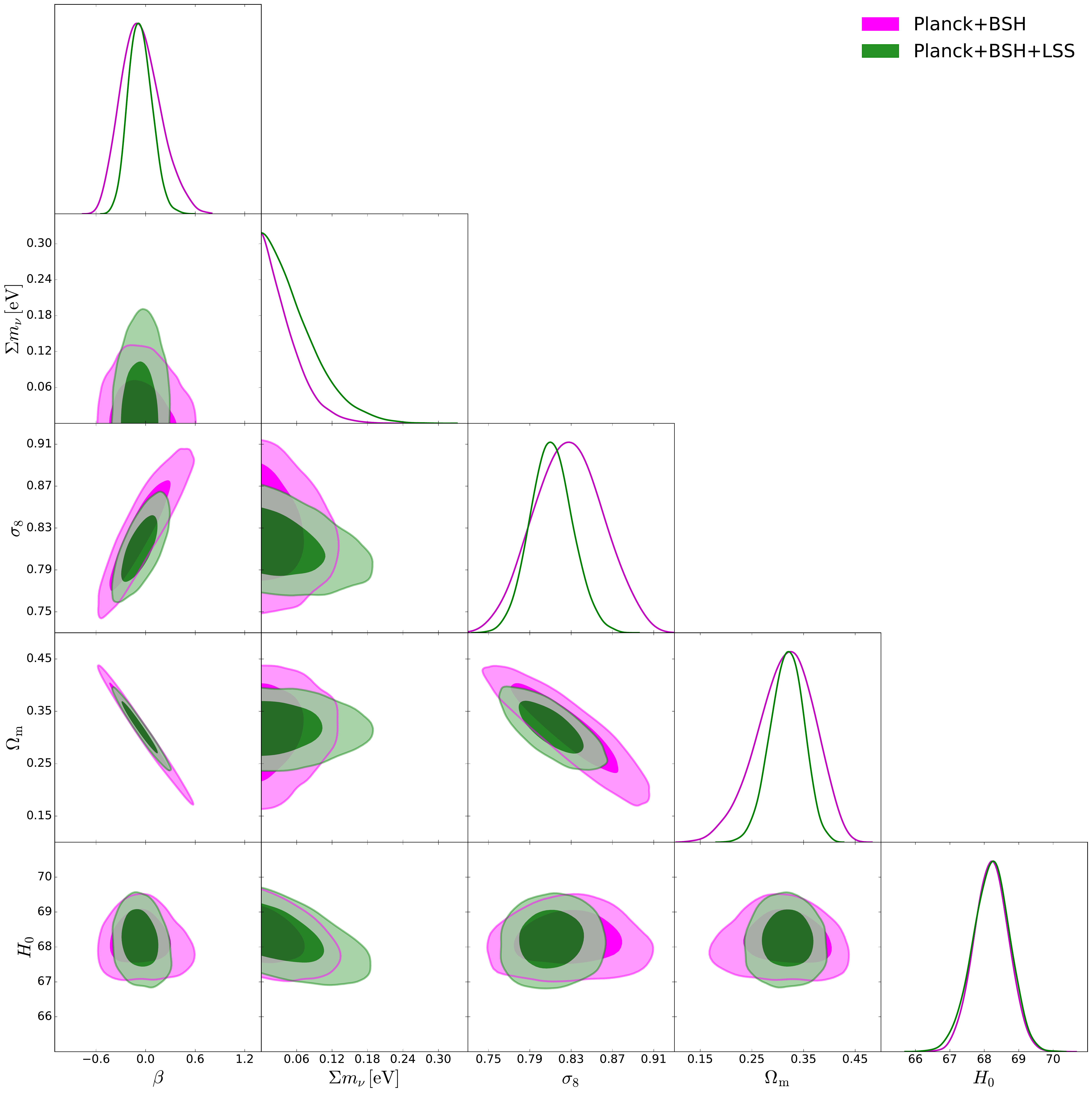}
\end{center}
\vspace{-8mm}
\caption{One-dimensional marginalized distributions and two-dimensional contours at $1\sigma$ and $2\sigma$ levels for parameters $\beta$, $\sum m_{\nu}$, $\Omega_{\rm m}$, and $H_{0}$ of the $Q=\beta H \rho_{\Lambda}$ model in the presence of $\sum m_{\nu}$ by using Planck+BSH and Planck+BSH+LSS.}
\label{fig6}
\end{figure*}

As described in Sec.~\ref{data}, RSD data can exert a special influence on the $Q=\beta H \rho_{\Lambda}$ model (see also \cite{Li:2014cee}). We employ the updated RSD data with reduced uncertainties from the DR12 in this situation. We also consider the weak lensing data to constrain  parameters of the $Q=\beta H \rho_{\Lambda}$ model. For Planck+BSH and Planck+BSH+LSS, detailed fitting results are given in Table~\ref{table3}. We obtain the constraints on $\beta$ for the $Q=\beta H  \rho_{\Lambda}$ model without and with $\sum m_\nu$. From Planck+BSH, we have $\beta=-0.10\pm0.24$ for the case without $\sum m_\nu$ and $\beta=-0.04^{+0.20}_{-0.30}$ for the case with $\sum m_\nu$; from Planck+BSH+LSS, we have $\beta=-0.07^{+0.12}_{-0.14}$ for the case without $\sum m_\nu$ and $\beta=-0.08^{+0.13}_{-0.15}$ for the case with $\sum m_\nu$. The right panel of Fig.~\ref{fig2} describes the one-dimensional posterior distributions for the coupling parameter $\beta$ of the $Q=\beta H \rho_{\Lambda}$ model from Planck+BSH and Planck+BSH+LSS. These results slightly favor a negative coupling constant, which means vacuum energy decays into cold dark matter. From Figs.~\ref{fig5} and \ref{fig6}, we can clearly see that $\beta=0$ is actually consistent with the current observational data no matter if we involve $\sum m_{\nu}$ or not in the $Q=\beta H \rho_{\Lambda}$ model.

For the constraints on the neutrino mass in the $Q=\beta H  \rho_{\Lambda}$ model, we obtain $\sum m_{\nu}<0.10$ eV from Planck+BSH and $\sum m_{\nu}<0.14$ eV from Planck+BSH+LSS. Usually, adding the LSS data can substantially reduce the errors of parameters, but here adding the LSS data leads to a larger mass limit. The reason is that a lower $\sigma_{8}$ is favored by the current LSS observations, while in this case $\sum m_{\nu}$ is anti-correlated with $\sigma_{8}$,\footnote{The small-scale matter power spectrum is suppressed by massive neutrinos. For very small scales (with $k\gg k_{\rm nr}$ and $k\gg k_{\rm eq}$), we have $P(k)^{f_\nu}/P(k)^{f_\nu=0}\simeq 1-8f_\nu$, for $f_\nu<0.7$, where $f_\nu\equiv \Omega_\nu/\Omega_m$~\cite{Hu:1997mj}. Thus, a larger $\sum m_{\nu}$ will lead to a smaller $\sigma_{8}$. For more details, see review article \cite{Lesgourgues:2006nd} and references therein.} and thus a larger mass limit is derived (see Fig.~\ref{fig6}). Obviously, the result of $\sum m_{\nu}<0.10$ eV offers a hint to exclude the inverted hierarchy in this model. Recall that, in the $\Lambda$CDM model, we obtain $\sum m_{\nu}<0.11$ eV from Planck+BSH. Thus, it seems that we are now on the edge of diagnosing the mass hierarchies by using the cosmological observations to weigh neutrinos.

We notice that, compared to the $\Lambda$CDM model, in the I$\Lambda$CDM2 model (with $Q=\beta H  \rho_{\Lambda}$) that has one more parameter, we derive a slightly tighter constraint on $\sum m_\nu$ from Planck+BSH. Actually, in Refs.~\cite{Zhang:2015uhk,Wang:2016tsz}, it has been shown that in dynamical dark energy models the constraints on $\sum m_\nu$ can become both looser and tighter, compared to $\Lambda$CDM. In the holographic dark energy model that has one more parameter with respect to $\Lambda$CDM, the constraint on $\sum m_\nu$ is found to be much tighter than that in the $\Lambda$CDM model \cite{Zhang:2015uhk,Wang:2016tsz}. But, in the $w$CDM and $w_0w_a$CDM models, the constraints are much looser \cite{Zhang:2015uhk,Wang:2016tsz,Zhao:2016ecj} (see also \cite{Zhang:2017rbg}). Thus, in this work, the slightly tighter constraint on $\sum m_\nu$ derived in the $Q=\beta H  \rho_{\Lambda}$ model is not surprising.\footnote{The Planck mission accurately measures the acoustic peaks and the observed angular size of acoustic scale $\theta_\ast=r_s/D_A$ is determined to a high precision (much better than 0.1\% precision at 1$\sigma$). This measurement could place tight constraints on some combinations of cosmological parameters that determine $r_s$ and $D_A$. Using the Planck data to do cosmological fit, the parameter combinations must be constrained to be close to a surface of constant $\theta_\ast$, and this surface depends on the models assumed. When new parameters are introduced, the degeneracy in the parameter space gives rise to consistent changes in parameters so that the ratio of sound horizon and angular diameter distance remains nearly constant. In the current case, the coupling $\beta$ changes the density evolution of cold dark matter and vacuum energy, which will lead to some effects on the parameters including $\sum m_\nu$ because they would also change to compensate. In this case, $\beta<0$ is slightly more favored (although $\beta=0$ is still located in the 1$\sigma$ region), and this may lead to the fact that a slightly tighter constraint on $\sum m_\nu$ is derived.   }

Smaller mass limits derived in this paper are also perhaps due to the strong tension between Hubble constant $H_{0}$ and other observational data. A method to relieve the tension is to consider the dark radiation, with one more parameter $N_{\rm eff}$. Taking the $Q=\beta H \rho_{\Lambda}$ model as an example, we find that the upper limit on the neutrino mass is changed from $\sum m_{\nu}<0.10$ eV to $\sum m_{\nu}<0.15$ eV, from Planck+BSH. This confirms that the constraint results on $\sum m_{\nu}$ in our work are affected more or less by the tension between Hubble constant ($H_{0}=73.00\pm1.75$ km s$^{-1}$ Mpc$^{-1}$) and other observational data. More further studies on this issue should be done.

\section{Conclusion}\label{Conclusion}

In this paper, we constrain the neutrino mass in a scenario of vacuum energy interacting with cold dark matter. We wish to see how the cosmological weighing of neutrinos is affected by the coupling between dark energy and dark matter. It is well known that the 6-parameter base $\Lambda$CDM cosmology is favored by the latest Planck data, and in this work we only consider its minimal extension in which we fix $w=-1$ and only introduce a coupling parameter $\beta$ (except for the neutrino mass $\sum m_\nu$). Since the vacuum energy interacts with cold dark matter, it is not a pure background any more, and so we should consider its perturbation evolution. This is rather difficult in the traditional linear perturbation theory. To treat the dark energy perturbations in this case, we employ the PPF framework for the IDE scenario, which can provide a good calculation scheme for our situation and give stable cosmological perturbations. Within this framework, the whole parameter space of the I$\Lambda$CDM scenario can be explored by using the current (and future) observational data. We use the latest Planck CMB data, combined with other observations, to constrain the models. The main data combinations are Planck+BSH and Planck+BSH+LSS. The BAO and RSD data from the BOSS DR12 are used in the analysis. Also, the latest local measurement of the Hubble constant (for which the uncertainty is reduced from 3.3\% to 2.4\%) is also used in our study. 

We consider two typical interaction forms of $Q=\beta H  \rho_{\rm c}$ and $Q=\beta H \rho_{\Lambda}$ in the I$\Lambda$CDM scenario. Compared to the standard $\Lambda$CDM model, there is only one more parameter, $\beta$. For the $Q=\beta H \rho_{\rm c}$ model, we find that $\beta>0$ (at more than $1\sigma$ level) from Planck+BSH, which indicates that cold dark matter decays into vacuum energy. For the $Q=\beta H \rho_{\Lambda}$ model, we find that $\beta=0$ is consistent with the current cosmological observations within the 1$\sigma$ range, which implies that this case is prone to be reduced to the standard $\Lambda$CDM model.

In the standard $\Lambda$CDM model, we obtain $\sum m_{\nu}<0.11$ eV ($2\sigma$) from Planck+BSH. This small upper limit is derived because we use the latest BAO DR12 data and Hubble constant ($H_{0}=73.00\pm1.75$ km s$^{-1}$ Mpc$^{-1}$). Compared with the BAO data from the DR11, the DR12 data can provide a 12\% change on the constraint of neutrino mass, which leads to a smaller upper bound. In addition, a strong tension between the $H_{0}$ and other astronomical observations is also a key reason to reduce the upper limit value of neutrino mass. But we still employ the measurement of $H_{0}=73.00\pm1.75$ km s$^{-1}$ Mpc$^{-1}$ in our analysis, because of its largely reduced uncertainty.

In the I$\Lambda$CDM model with $Q=\beta H \rho_{\rm c}$, we obtain $\sum m_{\nu}<0.20$ eV, from Planck+BSH. In the I$\Lambda$CDM model with $Q=\beta H \rho_{\Lambda}$, we obtain $\sum m_{\nu}<0.10$ eV from Planck+BSH and $\sum m_{\nu}<0.14$ eV from Planck+BSH+LSS. Finally, we take the $Q=\beta H \rho_{\Lambda}$ model as an example and introduce dark radiation to relieve the tension between Hubble constant $H_{0}$ and other observational data. We find that the upper limit on the neutrino mass is changed from $\sum m_{\nu}<0.10$ eV to $\sum m_{\nu}<0.15$ eV, from Planck+BSH. Therefore, though we are now on the edge of determining the mass hierarchy by using the cosmological weighing of neutrinos, we still must be very careful about the tension between observational data sets. Further deeper investigations are needed.

\acknowledgments
This work was supported by the National Natural Science Foundation of China (Grants No.~11522540 and No.~11690021), the Top-Notch Young Talents Program of China, and the Provincial Department of Education of Liaoning (Grant No.~L2012087).

\end{document}